\documentstyle[pra,twocolumn,aps,amsfonts,amssymb,epsfig]{revtex}

\newcommand{\bra}[1]{\langle{#1}|}
\newcommand{\ket}[1]{|{#1}\rangle}
\newcommand{\tr}{{\rm tr \thinspace}}
\def\ketc[#1]{\vert #1 \rangle}
\def\brac[#1]{\langle #1 \vert}
\def\sigjk{\sigma_j^{(k)}}
\newcommand{\expect}[1]{\langle{#1}\rangle}
\newcommand{\expectc}[1]{\langle{#1}\rangle^{\phantom{\dagger}}_{\!c}}

\draft \preprint{IQI preprint}
\title{Continuous quantum error correction via quantum feedback control}
\author{Charlene Ahn\thanks{\tt cahn@theory.caltech.edu},
Andrew C. Doherty\thanks{\tt dohertya@caltech.edu}, and Andrew J.
Landahl\thanks{\tt alandahl@theory.caltech.edu}}
\address{Institute for Quantum Information\\
California Institute of Technology\\
Pasadena, CA 91125, USA}
\date{\today}
\begin{document}
\maketitle
\begin{abstract}
We describe a protocol for continuously protecting \emph{unknown}
quantum states from decoherence that incorporates design
principles from both quantum error correction and quantum feedback
control. Our protocol uses continuous measurements and Hamiltonian
operations, which are weaker control tools than are typically
assumed for quantum error correction.  We develop a cost function
appropriate for unknown quantum states and use it to optimize our
state-estimate feedback.  Using Monte Carlo simulations, we study
our protocol for the three-qubit bit-flip code in detail and
demonstrate that it can improve the fidelity of quantum states
beyond what is achievable using quantum error correction when the
time between quantum error correction cycles is limited.
\end{abstract}

\section{Introduction}\label{sec:intro}

Long-lived coherent quantum states are essential for many quantum
information science applications including quantum cryptography
\cite{secure_QKD}, quantum computation \cite{nc,ph219_notes}, and
quantum teleportation \cite{bennett1992a}. Unfortunately, coherent
quantum states have extremely short lifetimes in realistic open
quantum systems due to strong decohering interactions with the
environment. Overcoming this decoherence is the chief hurdle faced
by experimenters studying quantum-limited systems.

Quantum error correction is a ``software solution'' to this
problem \cite{shor95,steane96}.  It works by redundantly encoding
quantum information across many quantum systems.  The key to this
approach is the use of measurements which reveal information about
which errors have occurred and not about the encoded data. This
feature is particularly useful for protecting the unknown quantum
states that appear frequently in the course of quantum
computations. The physical tools used in this approach are
projective von Neumman measurements that discretize errors onto a
finite set and fast unitary gates that restore corrupted data.
When combined with fault-tolerant techniques, and when all noise
sources are below a critical value known as the accuracy
threshold, quantum error correction enables quantum computations
of arbitrary length with arbitrarily small output error, or
so-called fault-tolerant quantum computation
\cite{shor96,kitaev96}.

Quantum feedback control is also sometimes used to combat
decoherence \cite{wiseman_pra94,goetsch1996a,tombesi1995a}. This
approach has the advantage of working well even when control tools
are limited.  The information about the quantum state fed into the
controller typically comes from continuous measurements and the
operations the controller applies in response are typically
bounded-strength Hamiltonians.  The performance of the feedback
may also be optimized relative to the resources that are
available. For example, one can design a quantum feedback control
scheme which minimizes the distance between a quantum state and
its target subject to the constraint that all available
controlling manipulations have bounded strengths
\cite{doherty-jacobs}.

The availability of quantum error correction, which can protect
unknown quantum states, and quantum feedback control, which uses
weak measurements and slow controls, suggests that there might be
a way to merge these approaches into a single technique with all
of these features.  Previous work to account for continuous time
using quantum error correction has focused on ``automatic''
recovery and has neglected the role of continuous measurement
\cite{brun,chuang-yamamoto,paz-zurek,aqec}. On the other hand,
previous work on quantum state protection using quantum feedback
control has focused on protocols for known states and has not
addressed the issue of protecting unknown quantum states
\cite{wang1,korotkov}, however see \cite{mabuchi-zoller96} for
related work.

The paper is organized as follows. In Sec.\ \ref{sec:qfc} we
review quantum feedback control and introduce the formalism of
stochastic master equations.  In Sec.\ \ref{sec:qec} we present
the three-qubit bit-flip code as a simple example of a quantum
error-correcting code and sketch the general theory using the
stabilizer formalism. In Sec.\ \ref{sec:qefc} we present our
protocol for quantum error feedback control and derive an optimal
non-Markovian feedback strategy for it.  In Sec.\
\ref{sec:bfc-sim} we use Monte Carlo simulations to demonstrate
this strategy's efficacy for the bit-flip code and compare it to
discrete quantum error correction when the time between quantum
error correction cycles is finite. Section \ref{sec:conclusion}
concludes.

\newpage

\section{Quantum Feedback Control}\label{sec:qfc}
\subsection{Open quantum systems}\label{sec:oqs}

To describe quantum feedback control, we first need to describe
uncontrolled open quantum system dynamics.  Let $\cal S$ be an
open quantum system weakly coupled to a reservoir $\cal R$ whose
self-correlation time is much shorter than both the time scale of
the system's dynamics and the time scale of the system-reservoir
interaction.  The Born-Markov approximation applies in this case
and enables us to write down a \emph{master equation}
\cite{carmichael} describing the induced dynamics in $\cal S$:
\begin{equation}
\label{eqn:me}
\dot{\rho} = -i\left[H,\,\rho\right] +
\sum_{\mu=1}^m {\cal{D}}[c_\mu]\,\rho.
\end{equation}
Here $\rho$ denotes the reduced density matrix for $\cal S$, $H$
its Hamiltonian, and $\cal{D}$ a decohering Lindblad superoperator
that takes a system-reservoir coupling operator (or \emph{jump
operator\/}) as an argument and acts on density matrices as
\begin{equation}
{\cal D}[c]\rho = c\rho c^\dagger - {1\over2}c^\dagger c\rho -
{1\over2}\rho c^\dagger c.
\end{equation}

One way to derive this master equation is to imagine that the
reservoir continuously measures the system but quickly forgets the
outcomes because of rapid thermalization.  The induced dynamics on
$\cal S$ therefore appear as an average over all possible
\emph{quantum trajectories} that could have been recorded by the
reservoir.

What kinds of measurements can the reservoir continuously perform?
The most general measurement quantum mechanics allows is a
positive operator-valued measure (POVM) $\{E_j\}$ acting on $\cal
S$. According to a theorem by Kraus \cite{kraus}, the  POVM
$\{E_j\}$ can always be decomposed as
\begin{equation}
\sum_i \Omega_{ij}^\dagger \Omega_{ij} = E_j,
\end{equation}
such that its stochastic action is $\rho\to\rho_j$ with
probability $p_j = \tr(\rho E_j)$, where
\begin{equation}
\rho_j = {1\over\tr(\rho E_j)}
         \sum_i \Omega_{ij} \rho \Omega_{ij}^\dagger.
\end{equation}
This POVM is called a \emph{strong measurement} when it can
generate finite state changes and a \emph{weak measurement} when
it cannot \cite{Lloyd_weak_feedback}.  POVMs that generate the
master equation (\ref{eqn:me}) involve infinitesimal changes of
state, and therefore are weak measurements.

One reservoir POVM that results in the master equation
(\ref{eqn:me}) is the continuous weak measurement with Kraus
operators
\begin{eqnarray}
\label{eqn:kraus1}
\Omega_0(dt) &=& 1 - \left(iH + {1\over2}\sum_{\mu=1}^m c_\mu^\dagger c_\mu\right)\,dt \\
\label{eqn:kraus2} \Omega_\mu(dt) &=& \sqrt{dt}\,c_\mu, \qquad \mu
= 1, \ldots, m.
\end{eqnarray}
Moreover, any POVM related to the one above via the unitary
rotation
\begin{equation}
\Omega'_\alpha = \sum_{\beta} U_{\alpha\beta} \Omega_{\beta}
\end{equation}
will generate the same master equation.  We call each distinct
POVM that generates the master equation when averaged over quantum
trajectories an \emph{unravelling} \cite{wiseman96} of the master
equation.

\subsection{Quantum feedback control}

The previous discussion of the master equation suggests a route
for feedback control. If we replace the reservoir with a device
that records the measurement current, then we could feed the
measurement record back into the system's dynamics by way of a
controller.  For example, the master equation (\ref{eqn:me}) with
$m=1$ can be unravelled into the \emph{stochastic master equation}
(SME) \cite{carmichael,fieldquad}
\begin{eqnarray}
\label{eqn:sme} d\rho_c(t) &=& -i\left[H,\,\rho_c(t)\right] dt \nonumber\\
                           & & + {\cal D}[c] \rho_c(t) dt
                               + {\cal H}[c] \rho_c(t) dW(t) \\
\label{eqn:current} dQ(t) &=& \expectc{c + c^\dagger} dt +
dW(t),
\end{eqnarray}
where $\rho_c$ is the conditioned density matrix, conditioned on
the outcomes of the measurement record $Q(t)$, the expectation
$\expectc{a}$ means $\tr(\rho_c a)$, $dW$ is a normally
distributed infinitesimal random variable with mean zero and
variance $dt$ (a \emph{Wiener increment} \cite{gardiner}), and
${\cal H}$ is a superoperator that takes a jump operator as an
argument and acts on density matrices as
\begin{eqnarray}
{\cal H}[c]\rho &=& c\rho + \rho c^\dagger
                   - \rho\, \tr[{c\rho + \rho c^\dagger}].
\end{eqnarray}

This sort of unravelling occurs, for example, when one performs a
continuous weak homodyne measurement of a field $c$ by first
mixing it with a classical local oscillator in a beamsplitter and
then measuring the output beams with photodetectors
\cite{fieldquad}. The stochastic model
(\ref{eqn:sme}--\ref{eqn:current}) is flexible enough to
incorporate other noise sources such as detector inefficiency,
dark counts, time delays, and finite measurement bandwidth
\cite{mabuchi_bandwidth}.

We can add feedback control by introducing a $Q(t)$-dependent
Hamiltonian to the dynamics of $\rho_c$.  There are two
well-studied ways of doing this.  The first, and simplest, is to
use Wiseman-Milburn feedback \cite{wiseman_pra94,fieldquad}, or
\emph{current feedback}, in which the feedback depends only on the
instantaneous measurement current $I_Q(t) = dQ(t)/dt$. For
example, adding the Hamiltonian $I_Q(t) F$ to the SME
(\ref{eqn:sme}) using current feedback leads to the dynamics
\cite{wiseman_pra94}
\begin{eqnarray}
\label{eqn:current_feedback}
d\rho_c(t) &=& -i\left[H,\,\rho_c(t)\right] dt \nonumber\\
           & & + {\cal D}[c] \rho_c(t) dt
               + {\cal H}[c] \rho_c(t) dW(t) \nonumber\\
           & & -i\left[F, c\rho_c(t) + \rho_c(t) c^\dagger\right] dt \nonumber\\
           & & + {\cal D}[F] \rho_c(t) dt
               -i\left[F, \rho_c(t) \right] dW \\
dQ(t) &=& \expectc{c + c^\dagger} dt + dW(t).
\end{eqnarray}

The second, and more general, way to add feedback is to modulate
the Hamiltonian by a functional of the entire measurement record.
An important class of this kind of feedback is \emph{estimate
feedback} \cite{doherty-jacobs}, in which feedback is a function
of the current conditioned state estimate $\rho_c$.  This kind of
feedback is of especial interest because of the quantum Bellman
theorem \cite{bellman}, which proves that the optimal feedback
strategy will be a function only of conditioned state expectation
values for a large class of physically reasonable cost functions.
An example of such an estimate feedback control law analogous to
the current feedback Hamiltonian used in
(\ref{eqn:current_feedback}) is to add the Hamiltonian $\expectc{
I_Q(t)} F=\expectc{c+c^\dagger} F$, which depends on what we
\emph{expect} the current $I_Q(t)$ should be given the previous
measurement history rather than its actual instantaneous value.
Adding this feedback to the SME (\ref{eqn:sme}) leads to the
dynamics
\begin{eqnarray}
d\rho_c(t) &=& -i\left[H,\,\rho_c(t)\right] dt \nonumber\\
           & & + {\cal D}[c] \rho_c(t) dt
               + {\cal H}[c] \rho_c(t) dW(t) \nonumber\\
           & & -i\expectc{I_Q} \left[F,\rho_c(t)\right] dt \\
dQ(t) &=& \expectc{c + c^\dagger} dt + dW(t).
\end{eqnarray}

\section{Quantum Error Correction}\label{sec:qec}

Although quantum feedback control has many merits, it has not been
used to protect unknown quantum states from noise. Quantum error
correction, however, is specifically designed to protect unknown
quantum states; for this reason it has been an essential
ingredient in the design of quantum computers
\cite{gott_thesis,knill-laflamme,ftqc}. The salient aspects of
quantum error correction can already be seen in the three-qubit
bit-flip code, even though it is not a fully quantum error
correcting code. For that reason, we shall introduce quantum error
correction with this example and discuss its generalization using
the stabilizer formalism.

\subsection{The bit-flip code}\label{sec:bf-stab}

The bit-flip code protects a single two-state quantum system, or
qubit, from bit-flipping errors by mapping it onto the state of
three qubits:
\begin{eqnarray}
\label{eqn:bfc1}
\ket{0} &\to& \ket{000} \equiv \ket{\bar{0}} \\
\label{eqn:bfc2}
\ket{1} &\to& \ket{111} \equiv \ket{\bar{1}}.
\end{eqnarray}
The states $\ket{\bar 0}$ and $\ket{\bar 1}$ are called the
\emph{basis states} for the code and the space spanned by them is
called the \emph{codespace}, whose elements are called
\emph{codewords}.

After the qubits are subjected to noise, quantum error correction
proceeds in two steps.  First, the parities of neighboring qubits
are projectively measured.  These are the observables\footnote{We
use the notation of \cite{gott_thesis} in which $X$, $Y$, and $Z$
denote the Pauli matrices $\sigma_x$, $\sigma_y$ and $\sigma_z$
respectively, and concatenation denotes a tensor product (e.g.,
$ZZI = \sigma_z \otimes \sigma_z \otimes I$).}
\begin{eqnarray}
\label{eqn:m1}
M_0 &=& ZZI \\
\label{eqn:m2}
M_1 &=& IZZ.
\end{eqnarray}
The \emph{error syndrome} is the pair of eigenvalues $(m_0, m_1)$
returned by this measurement.

Once the error syndrome is known, the second step is to apply one
of the following unitary operations conditioned on the error
syndrome:
\begin{eqnarray}
(-1, +1) &\to& XII \\
(-1, -1) &\to& IXI \\
(+1, -1) &\to& IIX \\
(+1, +1) &\to& III.
\end{eqnarray}

This procedure has two particularly appealing characteristics: the
error syndrome measurement does not distinguish between the
codewords, and the projective nature of the measurement
discretizes all possible quantum errors onto a finite set.  These
properties hold for general stabilizer codes as well.

If the bit-flipping errors arise from reservoir-induced
decoherence, then prior to quantum error correction the qubits
evolve via the master equation
\begin{equation}
\label{eqn:bit-flip_decoh} d\rho_{\mathrm noise} = \gamma ({\cal
D}[XII] + {\cal D}[IXI] + {\cal D}[IIX])\rho\,dt,
\end{equation}
where $\gamma dt$ is the probability of a bit-flip error on each
qubit per time interval $[t, t + dt]$. This master equation has
the solution
\begin{eqnarray}
  \label{eq:soln}
\lefteqn{\rho(t) =} \nonumber\\
          & & \phantom{+} a \left( t\right)\rho_{0} \nonumber \\
          & & + b \left( t\right)\left(XII\rho_{0}XII
                                     + IXI\rho_{0}IXI
                                     + IIX\rho_{0}IIX\right) \nonumber \\
          & & + c \left( t\right)\left(XXI\rho_{0}XXI
                                     + XIX\rho_{0}XIX
                                     + IXX\rho_{0}IIX\right) \nonumber \\
          & & + d \left( t\right) XXX\rho_{0}XXX,
\end{eqnarray}
where
\begin{eqnarray}
a(t) &=& \left(1+3e^{-2\gamma t}+3e^{-4\gamma t}+e^{-6\gamma t}\right)/8 \\
b(t) &=& \left(1+\phantom{3}e^{-2\gamma t}-\phantom{3}e^{-4\gamma t}-e^{-6\gamma t}\right)/8 \\
c(t) &=& \left(1-\phantom{3}e^{-2\gamma t}-\phantom{3}e^{-4\gamma t}+e^{-6\gamma t}\right)/8 \\
d(t) &=& \left(1-3e^{-2\gamma t}+3e^{-4\gamma t}-e^{-6\gamma t}\right)/8.
\end{eqnarray}

The functions $a(t)$--$d(t)$ express the probability that the
system is left in a state that can be reached by zero, one, two,
or three bit-flips from the initial state, respectively. After
quantum error correction is performed, single errors are
identified correctly but double and triple errors are not. As a
result, the recovered state, averaged over all possible
measurement syndromes, is
\begin{equation}
  \label{eq:corrected}
  \rho =\left( a \left( t\right) + b \left( t\right) \right)
  \rho _{0}+\left( c \left( t\right) + d \left( t\right)
  \right) XXX\rho _{0}XXX.
\end{equation}
The overlap of this state with the initial state depends on the
initial state, but is at least as large as when the initial state
is $\ket{\bar{0}}$; namely, it is at least as large as
\begin{eqnarray}
  \label{eqn:fidcorr}
  F_{\bar{3}} &=&\left( 2+3e^{-2\gamma t}-e^{-6\gamma t}\right) /4 \nonumber \\
     &\simeq& 1-3(\gamma t)^2.
\end{eqnarray}
Recalling that a single qubit subject to this decoherence has
error probability $p=\gamma t$, we see that, when applied
sufficiently often, the bit-flip code reduces the error
probability on each qubit from ${\cal O}(p)$ to ${\cal O}(p^2)$.

\subsection{Stabilizer formalism}\label{sec:stab}

The bit-flip code is one of many quantum error correcting codes
that can be described by the \emph{stabilizer formalism}
\cite{gott_thesis}. Let $\cal C$ be a $2^k$-dimensional subspace
of a $2^n$-dimensional $n$-qubit Hilbert space. Then the system
can be thought of as encoding $k$ qubits in $n$, where the
codewords are elements of $\cal C$. Let us further define the
\emph{Pauli group} to be $P_n = \{\pm 1, \pm i\}\otimes
\{I,X,Y,Z\}^{\otimes n}$, and let the \emph{weight} of an operator
in $P_n$ be the number of non-identity components it has when
written as a tensor product of operators in $P_1$. The
\emph{stabilizer} of $\cal C$, $S({\cal C})$, is the group of
operators which fix all codewords in $\cal C$. We call $\cal C$ an
$[[n,k,d]]$ \emph{stabilizer code} when ($a$) the $n-k$ generators
of $S({\cal C})$ form a subgroup of $P_n$ and ($b$) $d$ is the
smallest weight of an element in $P_n \setminus S({\cal C})$ that
commutes with every element in $S({\cal C})$.

In this general setting, quantum error correction proceeds in two
steps.  First, one projectively measures the stabilizer generators
to infer the error syndrome. Second, one applies a unitary
recovery operator conditioned on the error syndrome. The strong
measurement used in this procedure guarantees that all errors,
even unitary errors, are discretized onto a finite set. For this
reason we will sometimes refer to this procedure as discrete
quantum error correction. When the noise rate is low and when
correction is applied sufficiently often, this procedure reduces
the error probability from ${\cal O}(p)$ to ${\cal O}(p^2)$.

\section{Continuous quantum error correction via quantum feedback control}\label{sec:qefc}

In this section, we present a method for continuously protecting
an unknown quantum state using weak measurement, state estimation,
and Hamiltonian correction.  As in the previous section, we
introduce this method via the bit-flip code and then generalize.

\subsection{Bit-flip code: Theoretical model}\label{sec:bf-theory}

Suppose $\rho$ is subjected to bit-flipping decoherence as in
(\ref{eqn:bit-flip_decoh}); to protect against such decoherence,
we have seen that we can encode $\rho$ using the bit-flip code
(\ref{eqn:bfc1}--\ref{eqn:bfc2}).  Here we shall define a similar
protocol that operates continuously and uses only weak
measurements and slow corrections.

The first part of our protocol is to weakly measure the stabilizer
generators $ZZI$ and $IZZ$ for the bit-flip code, even though
these measurements will not completely collapse the errors.  To
localize the errors even further, we also measure the remaining
nontrivial stabilizer operator $ZIZ$.\footnote{The modest
improvement gained by this extra measurement is offset by an
unfavorable scaling in the number of extra measurements required
when applied to general $[[n, k, d]]$ codes having $2^{n-k}$
stabilizer elements and only $n-k$ generators.} The second part of
our protocol is to apply the slow Hamiltonian corrections $XII$,
$IXI$, and $IIX$ corresponding to the unitary corrections $XII$,
$IXI$, and $IIX$, with control parameters $\lambda_k$ that are to
be determined.  If we parameterize the measurement strength by
$\kappa$ and perform the measurements using the unravelling
(\ref{eqn:sme}--\ref{eqn:current}), the SME describing our
protocol is
\begin{eqnarray}
\label{eqn:bf-me}
d\rho_c &=& \phantom{+} \gamma ({\cal D}[XII]
                              + {\cal D}[IXI]
                              + {\cal D}[IIX])\rho_c dt \nonumber \\
        & & + \kappa ({\cal D}[ZZI]
                    + {\cal D}[IZZ]
                    + {\cal D}[ZIZ])\rho_c dt \nonumber\\
        & & + \sqrt{\kappa} ({\cal H}[ZZI] dW_1 + {\cal H}[IZZ] dW_2 \nonumber\\
        & & \phantom{+ \kappa(} + {\cal H}[ZIZ] dW_3)\rho_c \nonumber\\
        & & -  i [F, \rho_c] dt \\
\label{eqn:bfcurrent1}
dQ_1 &=& 2 \kappa \expectc{ZZI} dt + \sqrt{\kappa} dW_1 \\
\label{eqn:bfcurrent2}
dQ_2 &=& 2 \kappa \expectc{IZZ} dt + \sqrt{\kappa} dW_2 \\
\label{eqn:bfcurrent3}
dQ_3 &=& 2 \kappa \expectc{ZIZ} dt + \sqrt{\kappa} dW_3,
\end{eqnarray}
where
\begin{equation}
F = \lambda_1 XII + \lambda_2 IXI + \lambda_3 IIX
\end{equation}
is the feedback Hamiltonian having control parameters $\lambda_k$.

Following the logic of quantum error correction, it is
natural to choose the $\lambda_k$ to be functions of the error
syndrome. For example, the choice
\begin{eqnarray}
\label{eqn:oldfb}
\lambda_1 &=& {\lambda\over8}(1 - \expectc{ZZI})
                             (1 + \expectc{IZZ})
                             (1 - \expectc{ZIZ}) \nonumber \\
\lambda_2 &=& {\lambda\over8}(1 - \expectc{ZZI})
                             (1 - \expectc{IZZ})
                             (1 + \expectc{ZIZ}) \nonumber \\
\lambda_3 &=& {\lambda\over8}(1 + \expectc{ZZI})
                             (1 - \expectc{IZZ})
                             (1 - \expectc{ZIZ}),
\end{eqnarray}
where $\lambda$ is the maximum feedback strength that can be
applied, is reasonable\footnote{The factor of $1\over8$ is
included to limit the maximal strength of any parameter
$\lambda_k$ to $\lambda$.}: it acts trivially when the state is in
the codespace and applies a maximal correction when the state is
orthogonal to the codespace. Unfortunately this feedback is
sometimes harmful when it need not be. For example, when the
controller receives no measurement inputs (i.e., $\kappa = 0$), it
still adds an extra coherent evolution which, on average, will
drive the state of the system away from the state we wish to
protect.

This weakness of the feedback strategy suggests that we should
choose our feedback more carefully.  To do this, we introduce a
cost function describing how far away our state is from its target
and choose a control which minimizes this cost. The difficulty is
that our target is an \emph{unknown} quantum state. However, we
can choose the target to be the codespace, which we do know.  We
choose our cost function, therefore, to be the norm of the
component of the state outside the codespace. Since the codespace
projector is $\Pi_{\cal C} = \frac{1}{4}(III + ZZI + ZIZ + IZZ)$,
the cost function is $1-f$, where $f(\rho) = {\mathrm tr}(\rho
\Pi_{\cal C})$. Under the SME (\ref{eqn:bf-me}), the time
evolution of $f$ due to the feedback Hamiltonian $F$ is
\begin{eqnarray}
\label{eqn:dfdt}
\dot{f}_{fb} &=& \phantom{+} 2 \lambda_1 \expectc{YZI + YIZ} \nonumber\\
        & & +  2 \lambda_2 \expectc{ZYI + IYZ} \nonumber\\
        & & +  2 \lambda_3 \expectc{ZIY + IZY}.
\end{eqnarray}
Maximizing $\dot{f}_{fb}$ minimizes the cost, yielding the optimal
feedback coefficients
\begin{eqnarray}
\label{eqn:bf-fb}
\lambda_1 &=& \lambda\,{\mathrm sgn}\expectc{YZI + YIZ} \nonumber\\
\lambda_2 &=& \lambda\,{\mathrm sgn}\expectc{ZYI + IYZ} \nonumber\\
\lambda_3 &=& \lambda\,{\mathrm sgn}\expectc{ZIY + IZY},
\end{eqnarray}
where, again, $\lambda$ is the maximum feedback strength that can
be applied.

This feedback scheme is a \emph{bang-bang} control scheme, meaning
that the control parameters $\lambda_k$ are always at the maximum
or minimum value possible ($\lambda$ or $-\lambda$, respectively),
which is a typical control solution both classically
\cite{classical_bang-bang} and quantum mechanically
\cite{quantum_bang-bang}. In practice, the bang-bang optimal
controls (\ref{eqn:bf-fb}) can be approximated by a
bandwidth-limited sigmoid, such as a hyperbolic tangent function.

The control solution (\ref{eqn:bf-fb}) requires the controller to
integrate the SME (\ref{eqn:bf-me}) using the measurement currents
$Q_i(t)$ and the initial condition $\rho_c$.  However, typically
the initial state $\rho_c(0)$ will be unknown.  Fortunately the
calculation of the feedback (\ref{eqn:bf-fb}) does not depend on
where the initial condition is within the codespace, so the
controller may assume the maximally mixed initial condition
$\rho_e
=\frac{1}{2}(\ket{\bar{0}}\bra{\bar{0}}+\ket{\bar{1}}\bra{\bar{1}})$
for its calculations.  This property generalizes for a wide class
of stabilizer codes, as we prove in the appendix, and we
conjecture that this property holds for all stabilizer codes.

\subsection{Intuitive one-qubit picture}

Before generalizing our procedure, it is helpful to gain some
intuition about how it works by considering an even simpler
``code'': the spin-up state (i.e., $\ket{0}$) of a single qubit.
The stabilizer is $M_0 = Z$, the noise it protects against is bit
flips $X$, and the correction Hamiltonian is proportional to $X$.
The optimal feedback, by a similar analysis to that for the
bit-flip code, is $F = \lambda \, {\mathrm sgn}\langle Y \rangle_c
X$, and the resulting stochastic master equation can be rewritten
as a set of Bloch sphere equations as follows:

\begin{eqnarray}
\label{eqn:bloch-x}
d \expectc{X} &=& - 2 \kappa \expectc{X} dt
                 - 2 \sqrt{\kappa} \expectc{X} \expectc{Z} dW  \\
\label{eqn:bloch-y}
d \expectc{Y} &=& - 2 \gamma \expectc{Y} dt - 2 \kappa \expectc{Y}
                 - 2 \sqrt{\kappa} \expectc{Y} \expectc{Z} dW \nonumber \\
             && - 2 \lambda ({\mathrm sgn} \expectc{Y}) \expectc{Z} dt \\
\label{eqn:bloch-z}
d \expectc{Z} &=& - 2 \gamma \expectc{Z} dt
                 + 2\sqrt{\kappa} (1 - \expect{Z}^2_{\!c}) dW \nonumber \\
             && + 2 \lambda ({\mathrm sgn} \expectc{Y}) \expectc{Y} dt.
\end{eqnarray}

\begin{figure}
\begin{center}
\epsfig{file=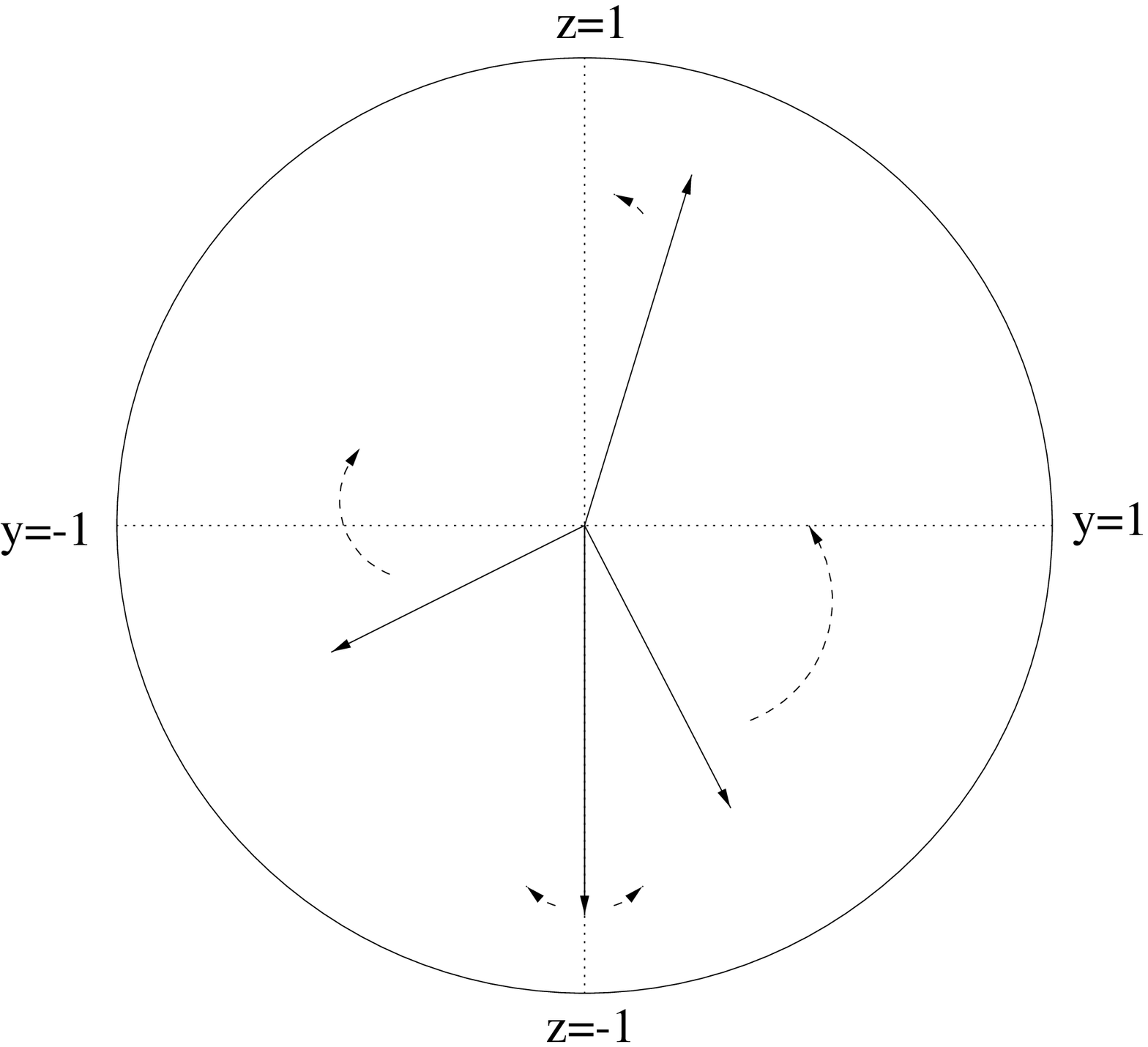,width=8.5cm}
\end{center}
\caption{Bloch sphere showing the action of our feedback scheme on
one qubit. Wherever the Bloch vector is in the $y$--$z$ plane, the
feedback forces it back to the spin-up state, which is the
codespace of this system. All the vectors shown lie, without loss
of generality, in the $x=0$ plane.} \label{fig:bloch}
\end{figure}

The Bloch vector representation $(\expect{X}, \expect{Y},
\expect{Z})$ \cite{ph219_notes} of the qubit provides a simple
geometric picture of how it evolves. Decoherence (the $\gamma$
term) shrinks the Bloch vector, measurement (the $\kappa$ terms)
lengthens the Bloch vector and moves it closer to the $z$-axis,
and correction (the $\lambda$ term) rotates the Bloch vector in
the $y$--$z$ plane. Fig.\ \ref{fig:bloch} depicts this evolution:
depending on whether the Bloch vector is in the hemisphere with
$\expect{Y} > 0$ or $\expect{Y} < 0$, the feedback will rotate the
vector as quickly as possible in such a way that it is always
moving towards the codespace (spin-up state). Note that if the
Bloch vector lies exactly on the $z$-axis with $\expect{Z} < 0$,
rotating it either way will move it towards the spin-up
state---the two directions are equivalent, and it suffices to
choose one of them arbitrarily.

\subsection{Feedback for a general code}

Our approach generalizes for a full $[[n, k, d]]$ quantum
error-correcting code, which can protect against depolarizing
noise \cite{ph219_notes} acting on each qubit independently. This
noise channel, unlike the bit-flip channel, generates a full range
of quantum errors---it applies either $X$, $Y$, or $Z$ to each
qubit equiprobably at a rate $\gamma$.  We weakly measure the
$n-k$ stabilizer generators $\{M_l\}$ with strength $\kappa$.  For
each syndrome $m$, we apply a slow Hamiltonian correction $F_m$
with control strength $\lambda_m$, the weight of each correction
being $d$ or less. The SME describing this process is
\begin{eqnarray}
\label{eqn:qeccme}
d\rho_c &=& \phantom{+} \gamma
\sum_{j=x,y,z}\sum_{i=1}^{n}
            ({\cal D}[\sigma^{(i)}_j])\rho_c dt
          + \kappa \sum_{l=1}^{n-k}{\cal D}[M_l]\rho_c dt\nonumber\\
        & & + \sqrt{\kappa} \sum_{l=1}^{n-k}{\cal H}[M_l] dW_j\rho_c
         -i \sum_{r=1}^R \lambda_r[F_r, \rho_c] dt.
\end{eqnarray}

The number of feedback terms $R$ needed will be less than or equal
to the number of errors the code corrects against.  The reason
that this equality is not strict is that quantum error correcting
codes can be \emph{degenerate}, meaning that there can exist
inequivalent errors that have the same effect on the state---a
purely quantum mechanical property \cite{gott_thesis}.

We optimize the $\lambda_r$ relative to a cost function equal to
the state's overlap with the codespace.   For a general stabilizer
code $\cal C$, the codespace projector is
\[
\Pi_{\cal C} = \frac{1}{2^{n-k}}\prod_{l=1}^{n-k}(I+M_l)
\]
and the rate of change of the codespace overlap due to feedback is
\[
\dot{f}_{fb} = -i\, {\mathrm tr} \sum_{r=0}^{n-k} \lambda_r
[\Pi_{\cal C}, F_r] \rho.
\]
Maximizing this overlap subject to a maximum feedback strength
$\lambda$ yields the feedback coefficients
\begin{equation}
\label{eqn:fb}
\lambda_r = \lambda\, {\mathrm sgn}\expectc{[\Pi_{\cal C} , F_r]}.
\end{equation}

This control solution, as for the bit-flip code, requires a
controller to compute the feedback (\ref{eqn:fb}).  A natural
question to ask is how the scaling of the classical computation
behaves. In the appendix we show that the evolution of
$(2^{n-k})^2$ parameters must be calculated in order to compute
the feedback for an $[[n,k,d]]$ code, which at first does not seem
promising. However, if one encodes $mk$ qubits using $m$ copies of
an $[[n,k,d]]$ code, as might well be the case for a quantum
memory, the SME (\ref{eqn:qeccme}) will not couple the dynamics of
the $m$ logical qubits; and, as in the bit-flip case, the initial
condition for the controller's integration can still be the
completely mixed state in the total codespace. Then the relevant
scaling for this system, the dependence on $m$, is linear: the
number of parameters is $m (2^{n-k})^2$.

\section{Simulation of the bit-flip code}\label{sec:bfc-sim}

In this section, we present the results of Monte Carlo simulations
of the implementation of the protocol described in Section
\ref{sec:qefc} for the bit-flip code.

\subsection{Simulation details}\label{sec:simdetail}

Because the bit-flip code feedback control scheme
(\ref{eqn:bf-me}--\ref{eqn:bfcurrent3}) uses a nonlinear feedback
Hamiltonian, numerical simulation is the most tractable route for
its study. To obtain $\rho_c(t)$, we directly integrated these
equations using a simple Euler integrator and a Gaussian random
number generator. We found stable convergent solutions when we
used a dimensionless time step $\gamma dt$ on the order of
$10^{-6}$ and averaged over $10^4$ quantum trajectories.  As a
benchmark, a typical run using these parameters took 2--8 hours on
a 400MHz Sun Ultra~2.  We found that more sophisticated Milstein
\cite{milstein} integrators converged more quickly but required
too steep a reduction in time step to achieve the same level of
stability. All of our simulations began in the state $\rho_c(0) =
\ket{\bar{0}}\bra{\bar{0}}$ because it is maximally damaged by
bit-flipping noise and therefore it yielded the most conservative
results.

We used two measures to assess the behavior of our bit-flip code
feedback control scheme. The first measure we used is the
\emph{codeword fidelity} $F_{cw}(t) = {\mathrm
tr}(\rho_c(0)\rho_c(t))$, the overlap of the state with the target
codeword.  This measure is appropriate when one cannot perform
strong measurements and fast unitary operations, a realistic
scenario for many physical systems. We compared $F_{cw}(t)$ to the
fidelities of one unprotected qubit $F_{1}(t) =
\frac{1}{2}(1+e^{-2\gamma t})$ and of three unprotected qubits
$F_{3}(t) = \left(F_{1}(t)\right)^3$.

The second measure we used is the \emph{correctable overlap}
\begin{equation}
 F_{corr}(t) = {\mathrm tr} (\rho_c(t) \Pi_{corr}),
\end{equation}
where
\begin{eqnarray}
\Pi_{corr} &=& \phantom{+} \rho_0 + XII\rho_0XII\nonumber\\
            &&+ IXI\rho_0IXI + IIX\rho_0IIX
\end{eqnarray}
is the projector onto the states that can be corrected back to the
original codeword by discrete quantum error correction applied
(once) at time $t$. This measure is appropriate when one can
perform strong measurements and fast unitary operations, but only
at discrete time intervals of length $t$.  We compared
$F_{corr}(t)$ to the fidelity $F_{\bar{3}}(t)$ obtained when,
instead of using our protocol up to time $t$, no correction was
performed until the final discrete quantum error correction at
time $t$. As we showed in equation (\ref{eqn:fidcorr}), the
expression for $F_{\bar{3}}(t)$ may be calculated analytically; it
is $F_{\bar{3}}(t) = \frac{1}{4}(2 + 3 e^{-2 \gamma t} - e^{-6
\gamma t}) \sim 1 - 3 \gamma^2 t^2$.

\subsection{Results}

\begin{figure}
\begin{center}
\epsfig{file=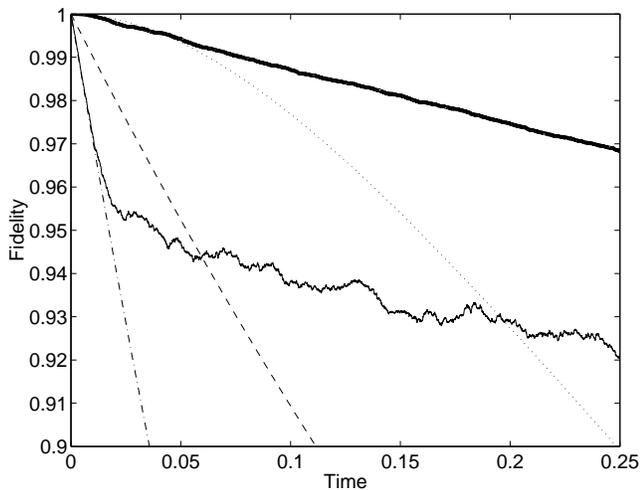,width=8.5cm}
\end{center}
\caption{Behavior of our protocol with optimized feedback
(\ref{eqn:bf-fb}) for parameters $\kappa/\gamma=64$,
$\lambda/\gamma=128$, averaged over $10^4$ quantum trajectories.
The analytical curves shown are as follows: the dashed line is the
fidelity of one decohering qubit, $F_1(t)$; the dashed-dotted line
is the fidelity of three decohering qubits, $F_3(t)$; and the
dotted line is the fidelity of an encoded qubit after one round of
discrete error correction, $F_{\bar{3}}(t)$. Our simulation
results are as follows: the solid line is the codeword fidelity
$F_{cw}(t)$, and the thick solid line is the correctable overlap
$F_{corr}(t)$.} \label{fig:fidelities}
\end{figure}

\begin{figure}
\begin{center}
\epsfig{file=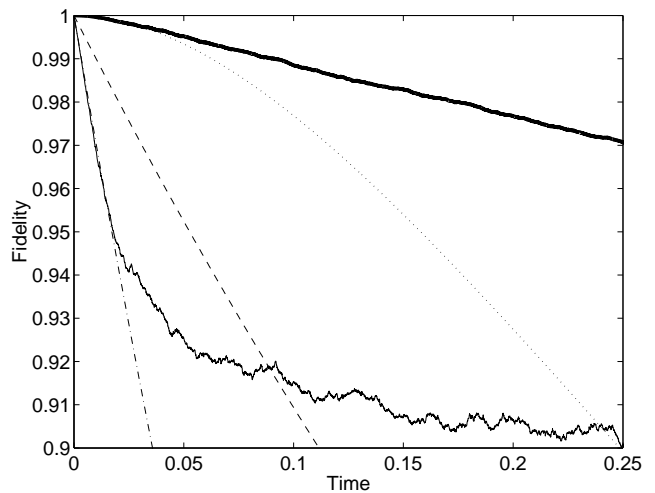,width=8.5cm}
\end{center}
\caption{Behavior of our protocol with non-optimized feedback
(\ref{eqn:oldfb}) for parameters $\kappa/\gamma=64$,
$\lambda/\gamma=128$, averaged over $10^4$ quantum trajectories.
As in Fig.\ \ref{fig:fidelities}, the dashed line is $F_1(t)$, the
dashed-dotted line is $F_3(t)$, the dotted line is
$F_{\bar{3}}(t)$, the solid line is $F_{cw}(t)$ and the thick
solid line is $F_{cw}(t)$. Note that this feedback is
qualitatively similar to that in Fig.\ \ref{fig:fidelities} but
does not perform as well.} \label{fig:old-feedback}
\end{figure}

We find that both our optimized estimate feedback scheme
(\ref{eqn:bf-fb}) and our heuristically-motivated feedback scheme
(\ref{eqn:oldfb}) effectively protect a qubit from bit-flip
decoherence. In Figs.\ \ref{fig:fidelities} and
\ref{fig:old-feedback} we show how these schemes behave for the
(scaled) measurement and feedback strengths $\kappa/\gamma=64$,
$\lambda/\gamma = 128$ when averaged over $10^4$ quantum
trajectories. Using our first measure, we see that at very short
times, both schemes have codeword fidelities $F_{cw}(t)$ that
follow the three-qubit fidelity $F_3(t)$ closely. For both
schemes, $F_{cw}(t)$ improves and surpasses the fidelity of a
single unprotected qubit $F_1(t)$. Indeed, perhaps the most
exciting feature of these figures is that eventually $F_{cw}(t)$
surpasses $F_{\bar{3}}(t)$, the fidelity achievable by discrete
quantum error correction applied at time $t$.  In other words, our
scheme alone outperforms discrete quantum error correction alone
if the time between corrections is sufficiently long.

Looking at our second measure in Figs.\ \ref{fig:fidelities} and
\ref{fig:old-feedback}, we see that $F_{corr}(t)$ is as good as or
surpasses $F_{\bar{3}}(t)$ almost everywhere.  For times even as
short as a tenth of a decoherence time, the effect of using our
protocol between discrete quantum error correction cycles is quite
noticeable. This improvement suggests that, even when one can
approximate discrete quantum error correction but only apply it
every so often, it pays to use our protocol in between
corrections.  Therefore, our protocol offers a means of improving
the fidelity of a quantum memory even after the system has been
isolated as well as possible and discrete quantum error correction
is applied as frequently as possible.

There is a small time range from $t\cong 0.01$ to $t\cong 0.05$
for the parameters used in Fig.\ \ref{fig:fidelities} in which
using our protocol before discrete quantum error correction
actually underperforms not doing anything before the correction.
Our simulations suggest that the reason for this narrow window of
deficiency is that, in the absence of our protocol, it is possible
to have two errors on a qubit (e.g., two bit flips) that cancel
each other out before discrete quantum error correction is
performed. In contrast, our protocol will immediately start to
correct for the first error before the second one happens, so we
lose the advantage of this sort of cancellation. This view is
supported by the fact that $F_{corr}(t)$ in our simulations always
lies above the fidelity line obtained by subtracting such
fortuitous cancellations from $F_{\bar{3}}(t)$. In any case, this
window can be made arbitrarily small and pushed arbitrarily close
to the beginning of our protocol by increasing the measurement
strength $\kappa$ and the feedback strength $\lambda$.

In Figs.\ \ref{fig:fidelities} and \ref{fig:old-feedback}, the
$F_{cw}(t)$ line is much more jagged than the $F_{corr}(t)$ line.
The jaggedness in both of these lines is due to statistical noise
in our simulation and is reduced when we average over more than
$10^4$ trajectories.  The reason for the reduced noise in the
$F_{corr}(t)$ line has to do with the properties of discrete
quantum error correction---on average, neighboring states get
corrected back to the same state by discrete quantum error
correction, so noise fluctuations become smoothed out.

The improvement our optimized estimate feedback protocol yields
beyond our heuristically-motivated feedback protocol is more
noticeable in $F_{cw}(t)$ than in $F_{corr}(t)$ as seen in Figs.\
\ref{fig:fidelities} and \ref{fig:old-feedback}.  Our optimized
protocol acts to minimize the distance between the current state
and the codespace, not between the current state and the space of
states correctable back to the original codeword, so this
observation is perhaps not surprising.  In fact, optimizing
feedback relative to $F_{corr}(t)$ is not even possible without
knowing the codeword being protected.  Nevertheless, our optimized
protocol does perform better, so henceforth we shall restrict our
to discussion to it.

\begin{figure}
\begin{center}
\epsfig{file=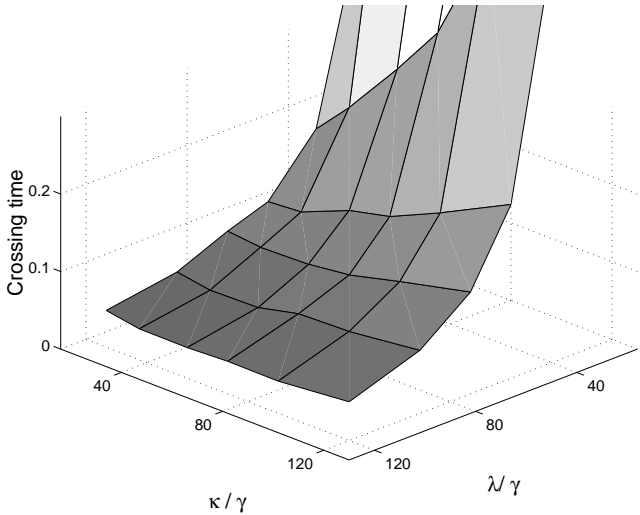,width=8.5cm}
\end{center}
\caption{Time $\tau$ at which $F_{cw}(\tau) = F_1(\tau)$ as a
function of measurement strength $\kappa/\gamma$ and feedback
strength $\lambda/\gamma$. This crossing time is the time after
which our optimized protocol improves the fidelity of a qubit
beyond what it would have been if it were left to itself.}
\label{fig:cw-1_crosstime}
\end{figure}

We investigated how our protocol behaved when the scaled
measurement strength $\kappa/\gamma$ and feedback strength
$\lambda/\gamma$ were varied using the two measures described in
Sec.\ \ref{sec:simdetail}.  Our first measure, the codeword
fidelity $F_{cw}(t)$, crosses the unprotected qubit fidelity
$F_1(t)$ at various times $\tau$ as depicted in Fig.\
\ref{fig:cw-1_crosstime}. This time is of interest because it is
the time after which our optimized protocol improves the fidelity
of a qubit beyond what it would have been if it were left to
itself.  Increasing the scaled feedback strength $\lambda/\gamma$
improves our scheme and reduces $\tau$, but the dependence on the
scaled measurement strength $\kappa/\gamma$ is not so obvious from
Fig.\ \ref{fig:cw-1_crosstime}.

\begin{figure}
\begin{center}
\epsfig{file=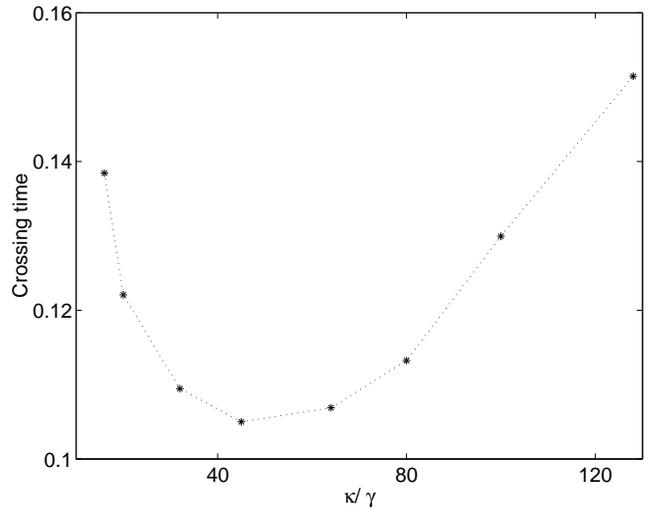,width=8.5cm}
\end{center}
\caption{Time $\tau$ at which $F_{cw}(\tau) = F_1(\tau)$ as a
function of measurement strength $\kappa/\gamma$, keeping
correction strength fixed at $\lambda/\gamma = 80$.}
\label{fig:lambda80}
\end{figure}

By looking at cross-sections of Fig.\ \ref{fig:cw-1_crosstime},
such as at $\lambda/\gamma=80$ as in Fig.\ \ref{fig:lambda80}, we
see that for a given scaled feedback strength $\lambda/\gamma$
there is a minimum crossing time $\tau$ as a function of
measurement strength $\kappa/\gamma$. In other words, there is an
optimal choice of measurement strength $\kappa/\gamma$. This
optimal choice arises because syndrome measurements, which
localize states near error subspaces, compete with Hamiltonian
correction operations, which coherently rotate states between the
nontrivial error subspaces to the trivial error subspace. This
phenomenon is a feature of our continuous-time protocol that is
not present in discrete quantum error correction; in the former,
measurement and correction are simultaneous, while in the latter,
measurement and correction are separate non-interfering processes.

\begin{figure}
\begin{center}
\epsfig{file=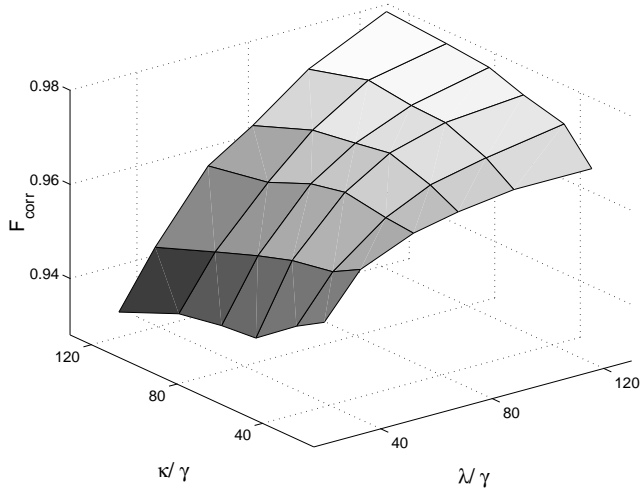,width=8.5cm}
\end{center}
\caption{$F_{corr}$ at $\gamma t=0.2$ as a function of measurement
strength $\kappa/\gamma$ and feedback strength $\lambda/\gamma$.
This quantity corresponds to the fidelity of a state given
continuous error-correction up to $\gamma t=0.2$, at which point
discrete error-correction is performed.} \label{fig:swqecc0.2}
\end{figure}

In order to study how our second measure, the correctable overlap
$F_{corr}(t)$, varies with $\kappa$ and $\lambda$, we found it
instructive to examine its behavior at a particular time.  In
Fig.\ \ref{fig:swqecc0.2} we plot $F_{corr}(t)$, evaluated at the
time $t=0.2/\gamma$, as a function of $\kappa$ and $\lambda$. As
we found with the crossing time $\tau$, increasing $\lambda$
always improves performance, but increasing $\kappa$ does not
because measurement can compete with correction. Since
$F_{\bar{3}}(0.2/\gamma) \cong 0.927$, for all the $\kappa$ and
$\lambda$ plotted in Fig.\ \ref{fig:swqecc0.2}, using our protocol
between discrete quantum error correction intervals of time
$0.2/\gamma$ improves the reliability of the encoded data.

\section{Conclusion}\label{sec:conclusion}

In many realistic quantum computing architectures, weak
measurements and Hamiltonian operations are likely to be the tools
available to protect quantum states from decoherence. Moreover,
even quantum systems in which strong measurements and fast
operations are well-approximated, such as ion traps
\cite{ion_traps}, it is likely that these operations will only be
possible at some maximum rate. Our protocol is able to
continuously protect unknown quantum states using only weak
measurements and Hamiltonian corrections and can improve the
fidelity of quantum states beyond rate-limited quantum error
correction.  In addition, because our protocol responds to the
entire measurement record and not to instantaneous measurement
results, it will not propagate errors badly and therefore has a
limited inherent fault-tolerance that ordinary quantum error
correction does not.

We expect that our protocol will be applicable to other
continuous-time quantum information processes, such as reliable
state preparation and fault-tolerant quantum computation.  We also
expect that our approach will work when different continuous-time
measurement tools are available, such as direct photodetection.
Finally, although current computing technology has limited our
simulation investigation to few-qubit versions of our protocol, we
are confident that many of the salient features we found in our
three-qubit bit-flip code protocol will persist when our protocol
is applied to larger codes.

\section{Acknowledgements}\label{sec:ack}
We happily acknowledge helpful discussions with many colleagues,
including Salman Habib, Kurt Jacobs, Hideo Mabuchi, Gerard
Milburn, John Preskill, Benjamin Rahn, and Howard Wiseman.  We are
especially grateful to Howard Wiseman for discussions on
optimizing our control scheme.  This work was supported in part by
the National Science Foundation under grant number EIA-0086038 and
by the Department of Energy under grant number DE-FG03-92-ER40701.
A.~L. acknowledges support from an IBM faculty partnership award
and C.~A. acknowledges support from an NSF fellowship.

\appendix

\section{Feedback based on the completely mixed state}

Even though our quantum error correction feedback control scheme
described in Section \ref{sec:qefc} does not distinguish between
codewords, it is not obvious that we do not need to know the
initial codeword to integrate its SME and calculate the relevant
expectation values. Since we are interested in protecting unknown
quantum states, this property is crucial to our scheme's success.
Fortunately, for a large class of stabilizer codes, the
computation of the feedback can be done by assuming the initial
state is the completely mixed codespace state $\rho_e =
\frac{1}{2^{n}}\prod_{l=1}^{n-k}(I+M_l)$, which we prove here.

We begin by defining the set $G$ for the $[[n, k, d]]$ code $\cal
C$ with stabilizer $S({\cal C})$ as
\begin{equation}
\label{eqn:mixedstateset} G = \left\{\alpha s \left|\,
                 \alpha \in P_n, s \in S({\cal C}),
                 [s,\alpha] = 0\ \mbox{iff}\ |\alpha|\ \mbox{is even}
              \right.
    \right\},
\end{equation}
where $|\alpha|$ denotes the weight of $\alpha$ as defined in
section \ref{sec:stab}.

It is also useful to define the normalizer $N(S)$ for the code as
the group of operators which commute with every element in
$S({\cal C})$. The elements of $ N(S) \setminus S$ can be thought
of as the \emph{encoded operations} for the code---they move one
codeword to another.

We shall rewrite the conditions we require for the computation of
the feedback to be insensitive to the initial codeword in terms of
the \emph{Pauli basis coefficients} $R_g(\rho)$ which we define as
follows. Let $g = \sigma_{i_1}\otimes...\otimes\sigma_{i_n}$, where
$i_1 \ldots i_n$ take on the values $x, y, z, I$ and $\sigma_I = I$. Then

\begin{equation}
R_{g}(\rho) \equiv {\mathrm
tr}(\rho g)/2^{n} = \langle g \rangle /2^n,
\end{equation}
 We can then formulate the problem in terms of proving
conditions on $G$ as follows:

\begin{enumerate}
    \item For every $R_g$ used in our feedback scheme, $g\in G$.
    \item For every $g \in G$ and every $\rho_1$ and $\rho_2$ in $\cal C$,
    $R_g(\rho_1) = R_g(\rho_2)$.
    \item Evolution under the SME couples members of the set $\{R_g | g\in G\}$
    only to each other.
\end{enumerate}

\textbf{Theorem} \emph{Let $\cal C$ be an $[[n, 1, 3]]$
\footnote{The restriction to $[[n,1,3]]$ codes is for simplicity
of analysis; the proof may be extended to larger codes. Note that
for an $[[n,1,3]]$ code, the $F_l$ in the master equation
(\ref{eqn:qeccme}) are all of the form $\sigjk$, where this
notation denotes the weight-one Pauli operator $\sigma_j$ acting
on qubit $k$.} stabilizer code whose stabilizer $S({\cal C})$ has
generators of only even weight and whose encoded operations set
$N(S)\setminus S$ has elements of only odd weight.\footnote{It is
possible that this restriction may be able to be relaxed; however,
it is sufficiently general that it holds for the most well-known
codes, including the bit-flip code, the five-bit code, the Steane
code, and the nine-bit Shor code. This condition also ensures that
$G$ is \emph{consistent}, i.e., if $\alpha_j s_k \in G$ and
$\alpha_j = \alpha_n s_m$, then $\alpha_n$ and $s_m s_k$ also
fulfill the conditions for $\alpha_n (s_m s_k)$ to be in $G$.}
Then the conditions 1--3 above are satisfied; consequently, our
scheme does not require knowledge of where the initial codeword
lies in $\cal C$.}

\textbf{Proof:}

In this proof, any variable of the form $\alpha_a$ is an arbitrary
element of $P_n$, and any variable of the form $s_a$ is an
arbitrary element of $S({\cal C})$.  We prove each of the
conditions listed above separately.

\emph{Condition 1:} By construction, $G$ contains all $M$ of the
form $M = s_i \sigma_j^{(k)}$, where $[s_i, \sigma_j^{(k)}] \neq
0$. These are precisely the operators used to compute the feedback
in (\ref{eqn:fb}) for a code encoding one qubit.

\emph{Condition 2:} Let $g = \alpha s \in G$ and let $\rho \in
{\mathcal C}$. We know either $\alpha \in S$, $\alpha \in
N(S)\setminus S$, or $\alpha \notin N(S)$. Suppose $\alpha \in S$.
Then $g \in S$ acts trivially on all states in the codespace, so
$R_{g} = 1/2^{n}{\mathrm tr}(\rho g) = 1/2^{n}$ for this case. Now
suppose $\alpha \in N(S)\setminus S$. Then $[\alpha, s] = 0$, and
since $\alpha s \in G$, $|\alpha|$ is even. But every element of
$N(S) \setminus S$ has odd weight by hypothesis, which is a
contradiction. Hence $\alpha$ cannot be in $N(S) \setminus S$.
Finally, suppose  $\alpha \notin N(S)$. Then there exists some $s'
\in S$ such that $[\alpha, s'] \neq 0$; let $s'$ be such an
element. Then for $\ket{\psi}, \ket{\phi} \in \cal C$,

\begin{eqnarray}
\bra{\psi} \alpha \ket{\phi} &=& \bra{\psi} \alpha s' \ket{\phi}
        = - \bra{\psi} s' \alpha \ket{\phi}\nonumber\\
        &=& - \bra{\psi} \alpha \ket{\phi} = 0.
\end{eqnarray}

Hence for this case $R_{g} = 1/2^{n}{\mathrm tr}(\rho \alpha s) =
0$. Note that these expressions for $R_g$ must be the same no
matter where $\rho$ is in the codespace; therefore, for every $g
\in G$ and $\rho_1, \rho_2 \in {\cal C}$, $R_g(\rho_1) =
R_g(\rho_2)$.

\emph{Condition 3:} We prove this by considering $dR_{M}$, where
$M\in G$: we will show that $dR_{M} = f(\{R_{N}|N\in G\})$ for
some real function $f$. Now, for any $M \in P_n$, $dR_{M} =
{\mathrm Tr}(d\rho \ M)$, where $d\rho$ is given by the master
equation (\ref{eqn:qeccme}), and we can show condition 3 for each
term of the master equation separately. First, substituting in the
master equation shows that any term of the form ${\cal D}[c]\rho
dt$ contributes either 0 or the simple exponential damping term
$-2 R_{M}$ to $dR_{M}$ if $M$ and $c$ commute or anticommute,
respectively.

As for the master equation term ${\cal H}[s_j] dW_j\rho$, by
writing the master equation in the Pauli basis we can see that
$R_{N}$ contributes to $dR_{M}$ through this term precisely when
$N s_j = M$ and $\{s_j, N\} \neq 0$. Here we know that $M\in G$,
so we may write $M = \alpha_k s_l$ (with the appropriate
restriction on $[\alpha_k, s_l]$ depending on the weight of
$\alpha_k$) . $N = \alpha_k s_l s_j = \alpha_k s_m$, so the
condition above that $[s_j, N] = 0$ becomes $[s_j, \alpha_k s_l
s_j] = (\alpha_k[s_j,s_l s_j] + [s_j, \alpha_k]s_l s_j)$
$\Rightarrow [s_j,\alpha_k] = 0$. Therefore, $[\alpha_k, s_m] =
s_l[\alpha_k, s_j] + [\alpha_k, s_l] s_j = [\alpha_k, s_l] s_j$
which is zero or not depending on the original weight of
$\alpha_k$. So if $M = \alpha_k s_l$ is such that $M \in G$, $N =
\alpha_k s_m$ must fulfill that same condition, implying that $N
\in G$ also.

Similarly, $R_{N}$ contributes to $dR_{M}$ through the master
equation term $[\sigma^{(k)}_j,\rho]$ when $N \sigjk = M$ and
$[\sigjk, N] \neq 0$. Now, $M\in G$ so $M = \alpha_l s_m$, again
with the appropriate restriction on $[\alpha_l, s_m]$ depending on
the weight of $\alpha_l$. Then $N = \sigjk \alpha_l s_m \equiv
\alpha_n s_m$, so the condition above that $\{\sigjk,N\} \neq 0$
becomes
\begin{eqnarray}
\label{eqn:Ncond}
\{ \sigjk , \sigjk \alpha_l s_m \}
  &=& \sigjk [\sigjk, \alpha_l] s_m  + \sigjk \alpha_l \{\sigjk, s_m\}\nonumber\\
  &=& \sigjk \{\sigjk, \alpha_l\} s_m -\sigjk \alpha_l [\sigjk, s_m]\nonumber\\
&=& 0.
\end{eqnarray}

We can now divide the analysis of this term
 into two cases. Case 1 occurs when $\sigjk \alpha_l$ has weight
$| \alpha_l |$, implying that $\{\alpha_l, \sigjk\} = 0$. Then
$\{\sigjk, \sigjk \alpha_l s_m \} = - \sigjk \alpha_l [\sigjk,
s_m] = 0$, which implies that $[s_m, \alpha_n] = [s_m, \sigjk]
\alpha_l + \sigjk [s_m, \alpha_l] = \sigjk [s_m, \alpha_l]$. So
$[s_m,\alpha_n] = 0$ just when $[s_m,\alpha_l] = 0$, which means
that $N\in G$ since $| \alpha_n | = | \alpha_l |$.

In Case 2, $\sigjk \alpha_l$ has weight $| \alpha_l \pm 1 |
\Rightarrow [\alpha_l, \sigjk] = 0$. Then (\ref{eqn:Ncond})
becomes $\{\sigjk, \sigjk \alpha_l s_m \} = \sigjk \alpha_l
\{\sigjk, s_m\}= 0$, which implies that $[s_m, \alpha_n] = \{s_m,
\sigjk\} \alpha_l + \sigjk \{s_m, \alpha_l\} = \sigjk \{s_m,
\alpha_l\}$. So $[s_m,\alpha_n] = 0$ just when $\{s_m,\alpha_l\} =
0$, which means that $N\in G$ since $| \alpha_n | = | \alpha_l \pm
1|$. $\square$

\bigskip

Thus we have shown the three conditions that all the $R$'s used to
compute the feedback are of the form $R_{N \in G}$; that for a
given $M \in G$, $R_M$ will be the same for any state in the
codespace; and that evolution via the master equation mixes the
$R$'s of the form $R_{N\in G}$ only with each other. Therefore, we
can conclude that taking the initial state to be \emph{any} state
in the codespace, including the true initial state and the
entirely mixed state, produces the same expression for the
feedback when the master equation is evolved conditioned on a
measurement record, and so we do not have to know the true initial
state to use our protocol.

Another consequence of using the completely mixed state for
feedback arises from the fact that doing so corresponds to
discarding information about the state of the system. Therefore,
this procedure should reduce the number of parameters needed to
compute the feedback. Unfortunately, this only leads to a modest
reduction in the number of parameters, which can be found by using
a simple counting argument. There are $2^n/2^k = 2^{n-k}$
different error subspaces, including the no-error (code) space,
and if we start with the completely mixed state in the codespace
we do not need to worry at all about any movement within any of
these spaces. We must only worry about which error space we are
actually in, along with coherences between these spaces, so we
find that $(2^{n-k})^2$ parameters are needed to describe the
system.

\end{document}